\documentclass[useAMS,usenatbib,usegraphicx,twocolumn]{mn2e}
\usepackage{amsmath,amssymb,amsbsy,amsfonts}
\usepackage[british]{babel}
\usepackage{mdwmath}
\usepackage{color}
\usepackage{url}
\usepackage{mathrsfs}
\usepackage{array}
\usepackage{graphicx}
\usepackage{tikz}
\usepackage{units}
\usepackage{relsize}
\usepackage{enumerate}
\usepackage[plainpages=false,
	    colorlinks=true,
            anchorcolor=blue,
            linkcolor=blue,
            citecolor=blue,
            bookmarks=false]{hyperref}
\usepackage{times}
\usepackage{pdflscape}

\newcommand{\eq}[1]{Eq.\,\eqref{#1}}
\newcommand{\eqs}[1]{Eqs.\,\eqref{#1}}
\newcommand{\cc}{\mathrm{c}}
\newcommand{\co}{\mathrm{const.}}
\newcommand{\ud}{\mathrm{d}}

\newcommand{\cn}{\ensuremath{\mathrm{cn}}}

\newcommand{\G}{\mathrm{G}}

\newcommand{\8}{_{\infty}}

\renewcommand{\theequation}{\thesection.\arabic{equation}}

\title[Relativistic wind accretion]
{Relativistic wind accretion onto a Schwarzschild black hole}
\author[E.~Tejeda \& A.~Aguayo-Ortiz]
{Emilio Tejeda$^{1,2}$\thanks{E-mail: emilio.tejeda@conacyt.mx, 
aaguayo@astro.unam.mx} and Alejandro Aguayo-Ortiz$^2$ \\
$^1$ CONACyT --  Instituto de F\'isica y Matem\'aticas, Universidad Michoacana 
de San Nicol\'as de Hidalgo, Edificio C-3, Ciudad Universitaria, 58040 
Morelia,\\ \hspace{0.15cm} Michoac\'an, Mexico\\
$^2$ Instituto de Astronom\'ia, Universidad Nacional Aut\'onoma de M\'exico, AP 
70-264, 04510 Ciudad de M\'exico, Mexico }

\pagerange{\pageref{firstpage}--\pageref{lastpage}}

\begin{document}

\maketitle

\label{firstpage}

\begin{abstract}
We present a novel analytic model of relativistic wind accretion onto a 
Schwarzschild black hole. This model constitutes a general relativistic 
extension of the classical model of wind accretion by Bondi, Hoyle and 
Lyttleton (BHL). As in BHL, this model is based on the assumptions of steady 
state, axisymmetry and ballistic motion. Analytic expressions are provided for 
the wind streamlines while simple numerical schemes are presented for 
calculating the corresponding accretion rate and density field. The resulting 
accretion rate is greater in the relativistic model when compared to the 
Newtonian BHL one. Indeed, it is two times greater for asymptotic wind speeds 
$v\8\ge0.4\,\cc$ and more than an order of magnitude greater for 
$v\8\ge0.8\,\cc$. We have compared this full relativistic model versus 
numerical simulations performed with the free GPL hydrodynamics code 
\textit{aztekas} (\url{aztekas.org} \copyright 2008 Sergio Mendoza \& Daniel 
Olvera and \copyright 2018 Alejandro Aguayo-Ortiz \& Sergio Mendoza) and found 
a good agreement for the streamlines in the upstream region of the flow and 
also, to within 10\%, for the corresponding accretion rates.
\end{abstract}


\section{Introduction}
\label{s1}

Accretion physics has become a basic tenet of astrophysics ever since it was 
recognized that the process of accretion, especially when compact objects are 
involved, is one of the most efficient mechanisms for converting rest mass 
energy into luminosity at work in our universe  \citep{king02}. Indeed, it has 
been established that a thin accretion disc around a non-rotating black hole 
can reprocess as much as 10 per cent of the accreted gas rest mass into 
electromagnetic radiation, while for a maximally rotating black hole this 
figure can reach up to 46 per cent~\citep[see e.g.][]{longair2011}. 

The simplest accretion scenario consists of the stationary, 
spherically-symmetric solution first discussed by \cite{bondi52}, where he 
considered an infinitely large homogeneous gas cloud steadily accreting onto a 
central gravitational object. The general relativistic extension of this model 
was developed by \cite{michel72} who took a Schwarzschild black hole as the 
central accretor. 

In the so-called wind accretion scenario, the spherical symmetry approximation 
is relaxed by considering a non-zero relative velocity between the central 
object and the accreted medium \citep[cf.][]{edgar04,romero14}. As it turns 
out, even after assuming a steady state and axial symmetry, the problem becomes 
sufficiently complex as not to admit a full analytic solution in general. In 
the pioneering work of \cite{hoylely} and \cite{hoyle} (BHL hereafter), the 
authors provided an analytic model for supersonic wind accretion by adopting 
the so-called ballistic approximation, i.e.~by neglecting pressure gradients 
and assuming that the fluid's dynamics is solely dictated by the central 
object's gravitational field. This approximation is well suited to describe 
highly supersonic flows given that, within this regime, incoming fluid elements 
cannot oppose pressure gradients readily, effectively following nearly 
free-fall trajectories.

Further analytic solutions to wind accretion problems have been found for a 
perfect fluid with a stiff equation of state in general relativity 
\citep{petrich88} and for the corresponding non-relativistic case of an 
incompressible fluid \citep{tejeda18}.

The problem of wind accretion has also been the focus of various numerical 
studies,  from both a Newtonian perspective 
\citep{hunt71,shima,ruffert,Elmellah2015,Elmellah2018} and in general 
relativity \citep{petrich89,font98a,zanotti11,LG13,GG15,CO16,CO17}. 

In this article we introduce a simple, analytic model for a supersonic wind 
accreting onto a non-rotating black hole (Schwarzschild spacetime). This model 
is a general relativistic extension of the BHL model and follows closely the 
methodology outlined in \cite{tejeda1}, \cite{tejeda2,tejeda3}. In that series 
of works, we presented an analytic model of the accretion flow of a rotating 
dust cloud infalling towards a central object, first in a Newtonian regime 
\citep{tejeda1} and then in general relativity for a Schwarzschild black hole 
\citep{tejeda2} and for a Kerr black hole \citep{tejeda3}. See \citet{schroven} 
for a recent  extension of this model considering the case of charged dust 
particles accreting onto a Kerr-Newman black hole.

The article is organised as follows. In Section~\ref{s2} we give a brief review 
of the BHL model. In Section~\ref{s3} we introduce the new relativistic wind 
model, and in Section~\ref{s4} we present the comparison against relativistic 
hydrodynamic numerical simulations performed with the free GPL hydrodynamics 
code \textit{aztekas} \citep{OM08,aguayo18}.\footnote{\url{aztekas.org} 
\copyright 2008 Sergio Mendoza \& Daniel Olvera and \copyright 2018 Alejandro 
Aguayo-Ortiz \& Sergio Mendoza.} Finally, in Section~\ref{s5} we summarize the 
work and present our conclusions. The Appendix \ref{sA} presents the results of 
the benchmark test of \textit{aztekas} against \citeauthor{michel72}'s analytic 
model, as well as self-convergence and various resolution tests relevant to the 
present study.

\section{Bondi-Hoyle-Lyttleton model}
\label{s2}

In this Section we give a brief overview of the BHL accretion model as its 
general relativistic extension constitutes the scope of the present work. The 
BHL model deals with a steady, supersonic wind accreting onto a gravitational 
object of mass $M$. Considering that the accretor is held fixed at the centre 
of coordinates, infinitely far away from the central object, the wind is 
homogeneous and characterized by a uniform density $\rho\8$ and wind speed 
$v\8$. 

\begin{figure}
\begin{center}
  \includegraphics[width=0.45\textwidth]{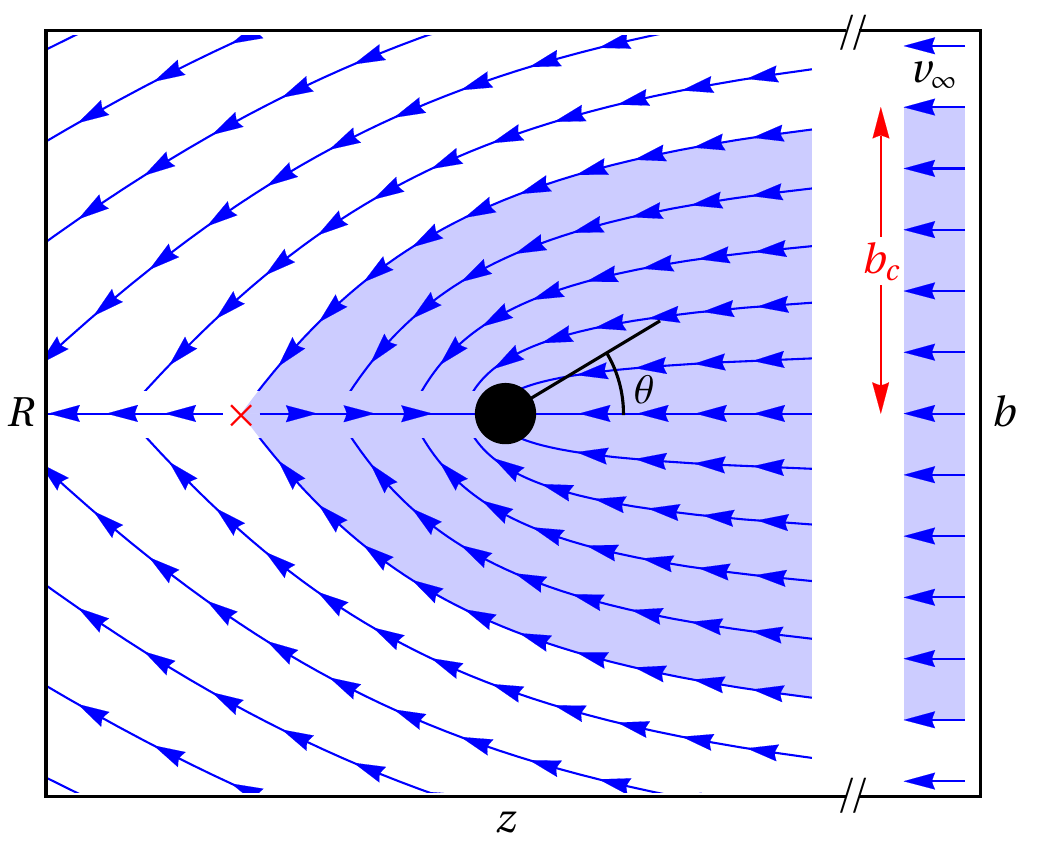}
\end{center}
\caption{Schematic representation of the BHL model. The incoming wind comes 
from an infinite distance to the right, corresponding to a polar angle $\theta 
= 0$. Under the ballistic approximation, fluid streamlines correspond to 
free-fall trajectories. At their arrival at the so-called accretion axis 
$\theta = \pi$, mirror symmetric streamlines collide against each other. The 
kinetic energy corresponding to the normal component of the velocity is 
efficiently thermalised and lost from the system. After this energy 
redistribution, streamlines with negative mechanical energy are left bound to 
the central object and constitute the accretion basin (shaded blue area). The 
stagnation point, i.e. the point along the accretion axis marking this 
transition, is marked with a red cross, and the corresponding streamline is 
characterized by the critical impact parameter $b_c$. The black circle shows 
the location of the central accretor. The axes correspond to the usual 
cylindrical coordinates $R = r\,\sin\theta$, $z = r\,\cos\theta$.} 
\label{fig1}
\end{figure}

Under the ballistic approximation, expressions for the streamlines, velocity 
field, and density for this model are given by \citep{bisnovatyi}
\begin{align}
r & = \frac{b^2v^2\8}{\G M(1-\cos\theta)+b\,v^2\8\sin\theta},
\label{e2.1}\\
\dot{r} & = -\sqrt{v^2\8+\frac{2\G M}{r}-\frac{b^2v^2\8}{r^2}},
\label{e2.2}\\
\dot{\theta} & = \frac{b\,v\8}{r^2},
\label{e2.3}\\
\rho & = \frac{\rho\8\,b^2}{r\,\sin\theta\left(2\,b-r\,\sin\theta\right)},
\label{e2.4}
\end{align}
where $b$ is the impact parameter of a given incoming fluid element. Note that 
we are taking here a reference frame in spherical coordinates such that the 
polar axis is aligned with the incoming wind direction. The wind comes 
asymptotically from the direction of the $\theta=0$ axis. Figure~\ref{fig1}
shows a schematic representation of the model setup and streamlines.

Given the uniform wind condition at infinity, all of the incoming trajectories 
have initially a common specific total energy $E = v^2\8/2$. In other words, 
the wind fluid elements follow energetically unbound (hyperbolic) trajectories. 
From \eqs{e2.1} and \eqref{e2.2} we see that the incoming trajectories reach 
the downstream axis at $r(\pi)=b^2v^2\8/(2\G M)$ with radial velocity 
$\dot{r}=v\8$. It is expected that streamlines coming from mirror-reflected 
points with respect to the symmetry axis will collide with one another along 
this axis (see Figure~\ref{fig1}). Following this collision, the BHL model 
envisages that the component of the velocity perpendicular to the axis is 
instantly transformed into thermal energy and, eventually, lost as 
radiation.\footnote{More precisely, in the \cite{hoylely} model the fluid 
streamlines are assumed to focus downstream onto the symmetry axis leading to 
an infinite density there. The accretion then proceeds along this line. 
Instead, \cite{hoyle} envisioned an accretion flow that spreads out onto a 
finite density region that they referred to as accretion column. In this work 
we are adopting the former description.} After this loss of kinetic energy, 
each fluid element is left with a new specific total energy
\begin{equation}
E' = \frac{1}{2}\dot{r}(\pi)^2 - \frac{\G M}{r(\pi)} = \frac{1}{2}v^2\8 - 
\frac{2(\G M)^2}{b^2v^2\8}.
\label{e2.5}
\end{equation}
By equating this energy to zero, the following critical value for the impact 
parameter is found
\begin{equation}
b_c = \frac{2\G M}{v^2\8},
\label{e2.6}
\end{equation}
such that any fluid element following a streamline with $b<b_c$ is 
energetically bound to the central object and, hence, eventually accreted. On 
the other hand, any fluid element following a streamline with $b>b_c$ is 
energetically unbound to the central object and, therefore, ultimately escapes 
to infinity. Following this argument, and accounting for all of the material 
within the cylinder of radius $b_c$, the total accretion rate onto the central 
object in the BHL model is given by
\begin{equation}
\dot{M}_{\mathrm{BHL}} = \pi\,b_c^2\,\rho\8\,v\8 = 4\,\pi\,\rho\8
\frac{(\G M)^2}{v^3\8}.
\label{e2.7}
\end{equation}
Subsequent numerical studies of this problem have found stationary accretion 
rates that agree remarkably well with those predicted by the simple BHL model 
\citep[see~e.g.][]{hunt71}. The resulting flows obtained in full-hydrodynamic 
simulations of supersonic wind accretion show the development a bow shock 
around the central accretor at which the incoming wind streamlines abruptly 
decelerate and become subsonic. Clearly, the simple analytic description of 
the streamlines provided by the BHL model is no longer valid inside the bow 
shock, nevertheless, it provides a qualitatively good description of the 
streamlines in the supersonic, upwind region.

\setcounter{equation}{0}
\section{Relativistic wind model}
\label{s3}

In this section we present the extension of the BHL wind model to the case in 
which the central accretor is a non-rotating black hole of mass $M$. We shall 
assume that the mass-energy content of the accreting gas is negligible as 
compared to the mass of the central black hole and, thus, that the overall 
spacetime metric corresponds to the Schwarzschild solution. For constructing 
this model we are closely following the methodology described in 
\cite{tejeda2}. For the remainder of this work we adopt a geometrised system of 
units for which $\G=\cc=1$.

\subsection{Velocity field}

Adopting the ballistic approximation for test particles in general relativity 
amounts to describe the incoming streamlines as geodesic trajectories. Thanks 
to the symmetries of Schwarzschild spacetime, the trajectory of a test particle 
is restricted to a plane and governed by the equations of motion \citep{novikov}
\begin{gather}
\frac{\ud t}{\ud \tau} = \mathcal{E} \left( 1 -\frac{2M}{r} \right)^{-1},
\label {e3.1} \\
\frac{\ud r}{\ud \tau} = -\left[\mathcal{E}^2 
-\left(1-\frac{2M}{r}\right)\left(1+\frac{h^2}{r^2}\right) \right]^{1/2},
\label {e3.2} \\
\frac{\ud \theta}{\ud \tau} = \frac{h}{r^2},
\label {e3.3} 
\end{gather}
where $\mathcal{E}$ is the conserved specific (relativistic) total energy and 
$h$ is the conserved specific angular momentum. Assuming a uniform wind velocity
at infinity $v\8$, these conserved quantities for a given streamline with 
impact parameter $b$ are given by
\begin{gather}
 \mathcal{E} = \frac{1}{\sqrt{1-v^2_\infty}} = \gamma\8 , 
 \label {e3.4} \\ 
 h = \frac{b\,v\8}{\sqrt{1-v^2_\infty}} = b\,V_\infty ,
 \label {e3.5} 
\end{gather}
where $\gamma\8$ is the wind's Lorentz factor at infinity and where we have 
introduced the shorthand notation $V_\infty = \gamma\8\,v\8$.

In order for the velocity field in \eqs{e3.1}-\eqref{e3.3} to be useful in 
practice, we need to provide an expression for the streamlines of the form 
$r(\theta,b)$ as we discuss in the next subsection.

\subsection{Streamlines}

An expression for the streamlines can be obtained by combining \eqs{e3.2} and 
\eqref{e3.3} as 
\begin{equation}
 \frac{\ud r}{ \ud \theta} = \frac{ \sqrt{\mathcal{R}(r) } }{h},
 \label{e3.6} 
\end{equation}
where 
\begin{equation}
 \mathcal{R}(r) = r\left[V^2_\infty \,r^3 + 2Mr^2 - b^2 V^2_\infty 
 (r- 2M)\right].
 \label{e3.7} 
\end{equation}

As discussed in detail in \cite{tejeda2}, \eq{e3.6} can be solved in terms of 
elliptic integrals.  For the problem at hand, we need to distinguish between 
two types of trajectories: 1) unbound trajectories that reach a minimum 
distance in their descent towards the central object before turning back to 
infinity and 2) trapped trajectories that plunge onto the black hole's event 
horizon (located at $r=2M$). Specifically, streamlines with $b\ge b_0$ belong 
to type 1) while those with $b<b_0$ belong to type 2) where
\begin{equation}
 b_0 = \frac{M\sqrt{ 27 \,V^4_\infty  + 18 \,V^2_\infty + \gamma_\infty \left(1 
+ 9\,V^2_\infty \right)^{3/2} -1 }}{\sqrt{2}\,V^2_\infty } .
 \label{e3.8}
\end{equation}

In case 1), the polynomial in \eq{e3.7} has three non-trivial real roots. For 
the particular boundary condition that we have adopted here, it can be proved 
that one of these roots is negative and the other two are positive. We call 
them $r_1$, $r_2$ and $r_3$, such that $r_1 < 0 < r_2<r_3$. These roots 
can be explicitly given in terms of the constants of motion, see 
e.g.~\cite{tejeda2}. In terms of these roots, the equation for the streamlines 
is given by 
\begin{gather}
r = \frac{ r_{1} (r_{3}-r_{2})-r_{2}(r_{3}-r_{1})
\cn^2 \left(\xi,\,k\right) }
{ r_{3}-r_{2}-(r_{3}-r_{1})\cn^2\left(\xi,\,k\right) },
 \label{e3.9} \\
\xi = \frac{\sqrt{r_{3}(r_{3}-r_{2})}}{2\,b}
( \theta + \theta\8), 
 \label{e3.10}
\end{gather}
where $\cn\left(\xi,\,k\right)$ is a Jacobi elliptic function with modulus 
\citep{lawden}
\begin{equation}
 k = \sqrt{\frac{r_2(r_{3}-r_{1})}{r_3(r_{2}-r_{1})}},
 \label{e3.11}
\end{equation}
and 
\begin{equation}
 \theta\8 = \frac{2\,b}{\sqrt{r_{3}(r_{3}-r_{2})}}\,
\cn^{-1}\sqrt{\frac{r_3-r_2}{r_3-r_1}},
 \label{e3.12}
\end{equation}
is the polar phase setting $\theta=0$ as the incoming wind direction, 
i.e.~$r(\theta=0)\rightarrow\infty$.

In case 2) the polynomial in \eq{e3.7} has as roots a negative real number 
$r_1$ and a complex conjugate pair $r_2=r_3^*$. In this case the equation 
for the streamlines can be written as
\begin{gather}
 r = \frac{ \delta\, r_1[1-\cn (\tilde\xi,\,\tilde k)]}
 { \delta-\eta - ( \eta +\delta) \cn (\tilde\xi,\,\tilde k) },
 \label{e3.13} \\
\tilde\xi =  \frac{\sqrt{ \eta\,\delta  }}{b}\, (\theta + \tilde\theta\8),
 \label{e3.14}
\end{gather}
where
\begin{gather}
 \eta = \sqrt{(r_2 - r_1)(r_3 - r_1)},
 \label{e3.15} \\
\delta = \sqrt{r_2 r_3},
\label{e3.16}\\
\tilde k^2 =  \frac{( \eta + \delta )^2 - r_1^2 }{ 4\, \eta\,\delta },
 \label{e3.17}
\end{gather}
and
\begin{equation}
 \tilde\theta\8 = \frac{b}{ \sqrt{ \eta\,\delta } }\,
\cn^{-1} \left[\frac{ \delta - \eta }
{ \eta+\delta },\,\tilde k\right] .
\label{e3.18}
\end{equation}

Equations \eqref{e3.9} and \eqref{e3.13} constitute the required expressions 
for the streamlines. As demonstrated by~\cite{tejeda2}, these expressions for 
the streamlines reduce to the usual Newtonian conic-sections in the 
non-relativistic limit, i.e.~for $v\8\ll1$ \eq{e3.9} reduces to \eq{e2.1}, 
while the velocity field in \eqs{e3.2} and \eqref{e3.3} reduce to \eqs{e2.2} 
and \eqref{e2.3}, respectively.
 
\subsection{Density field}

For calculating the density field we follow here a similar strategy as in 
\cite{tejeda2,tejeda3}. We start from the continuity equation\footnote{Here and 
in what follows Greek indices run over spacetime components, Latin indices run 
over spatial components only, and Einstein's summation convention over repeated 
indices is adopted.}
\begin{equation}
\nabla_\mu(\rho\,U^\mu)=0, 
 \label{e3.19}
\end{equation}
where $U^\mu = \ud x^\mu/\ud \tau$ is the four-velocity, $\rho$ is the rest 
mass density and $\nabla_\mu$ stands for the covariant derivative.

\begin{figure}
\begin{center}
  \includegraphics[width=0.48\textwidth]{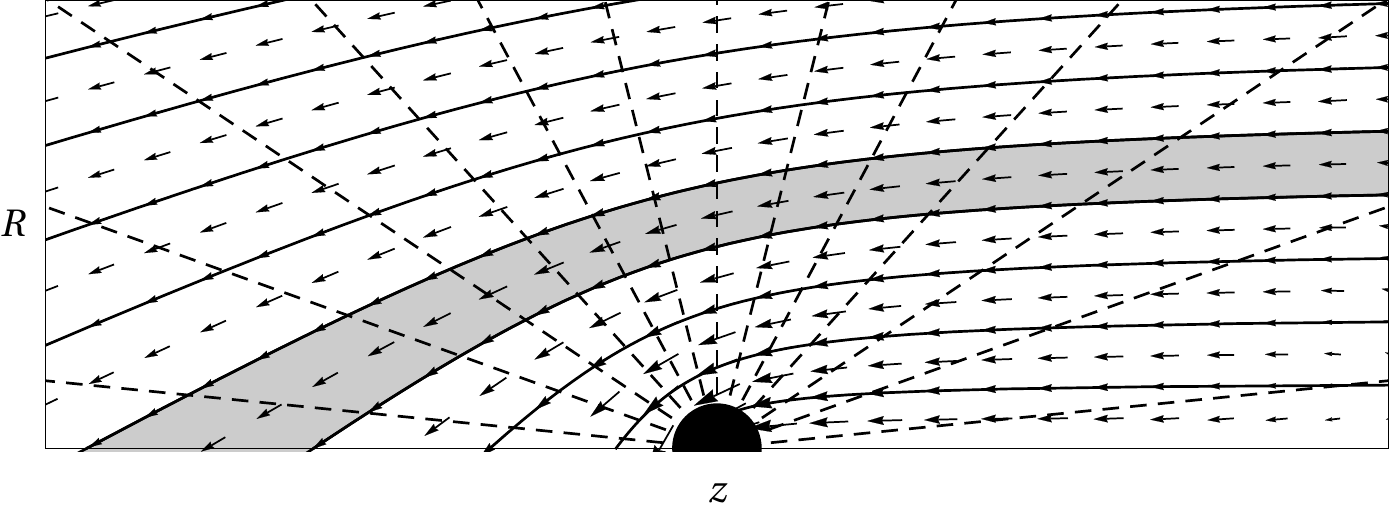}
\end{center}
\caption{Schematic illustration of the grid used for numerically evaluating the 
partial derivative involved in the calculation of the density field 
(Eq.~\ref{e3.23}). The shaded area represents one of the streamline tubes used 
as integration volume for calculating the density field in \eq{e3.20}.}
 \label{fig2}
\end{figure}

Writing \eq{e3.19} for a Schwarzschild spacetime and adopting the stationary 
condition results in
\begin{equation}
 \frac{\partial}{\partial x^i}(r^2\sin\theta\,\rho\,U^i) = 0.
\label{e3.20}
\end{equation}
Let us now integrate \eq{e3.20} over the spatial volume element delimited by a 
sufficiently small set of streamlines that start from an area element 
\mbox{$S\8=2\pi\,b\,\ud b$} located at infinity and that end at the 
intersection with the conic surface defined by $\theta=\co$, i.e.~
\mbox{$S_\theta=2\pi\,r\,\sin\theta\,\ud r$}, as shown in Figure~\ref{fig2}. By 
construction, fluid elements enter the integration volume only across $S\8$ and 
leave across $S_\theta$. Therefore, by means of Gauss's theorem, it follows that
\begin{equation}
 \rho\,r\,U^\theta\,S_\theta = \rho\8\,\gamma\8\,v\8\,S\8,
 \label{e3.21}
\end{equation}
from where
\begin{equation}
 \rho\,r^2 U^\theta\sin\theta\,\ud r = \rho\8\,\gamma\8\,v\8\,b\,\ud b.
 \label{e3.22}
\end{equation}
Finally, substituting $U^\theta$ from \eq{e3.3} into \eq{e3.22} results in
\begin{equation}
 \rho = \frac{\rho\8}{\sin\theta}\left(\frac{\partial r}{\partial 
b}\right)^{-1}.
 \label{e3.23}
\end{equation}

Calculating explicitly the partial derivative in \eq{e3.23} is something 
trivial to do in the Newtonian case. Indeed, this calculation is the step 
leading from \eq{e2.1} to \eq{e2.4}. However, this same calculation in the 
relativistic case represents a complex procedure involving the derivative of an 
elliptic function with respect to its argument and modulus. Following 
\cite{tejeda2}, we do not attempt here to calculate this derivative 
analytically but rather use a finite difference scheme to compute it 
numerically. A suitable grid for performing this calculation can be constructed 
as follows: Start from a collection of streamlines separated by uniform 
intervals of $\Delta b$ at infinity. Follow these streamlines from $\theta = 0$ 
to $\theta = \pi$ storing the different values of $r$ at uniform steps of 
$\Delta \theta$. Use these grid values for estimating $\partial r/\partial b$ 
as a finite difference of the radial coordinate between neighboring 
streamlines. Such a grid is schematically represented in Figure~\ref{fig2}. 

\begin{figure}
\begin{center}
  \includegraphics[width=0.48\textwidth]{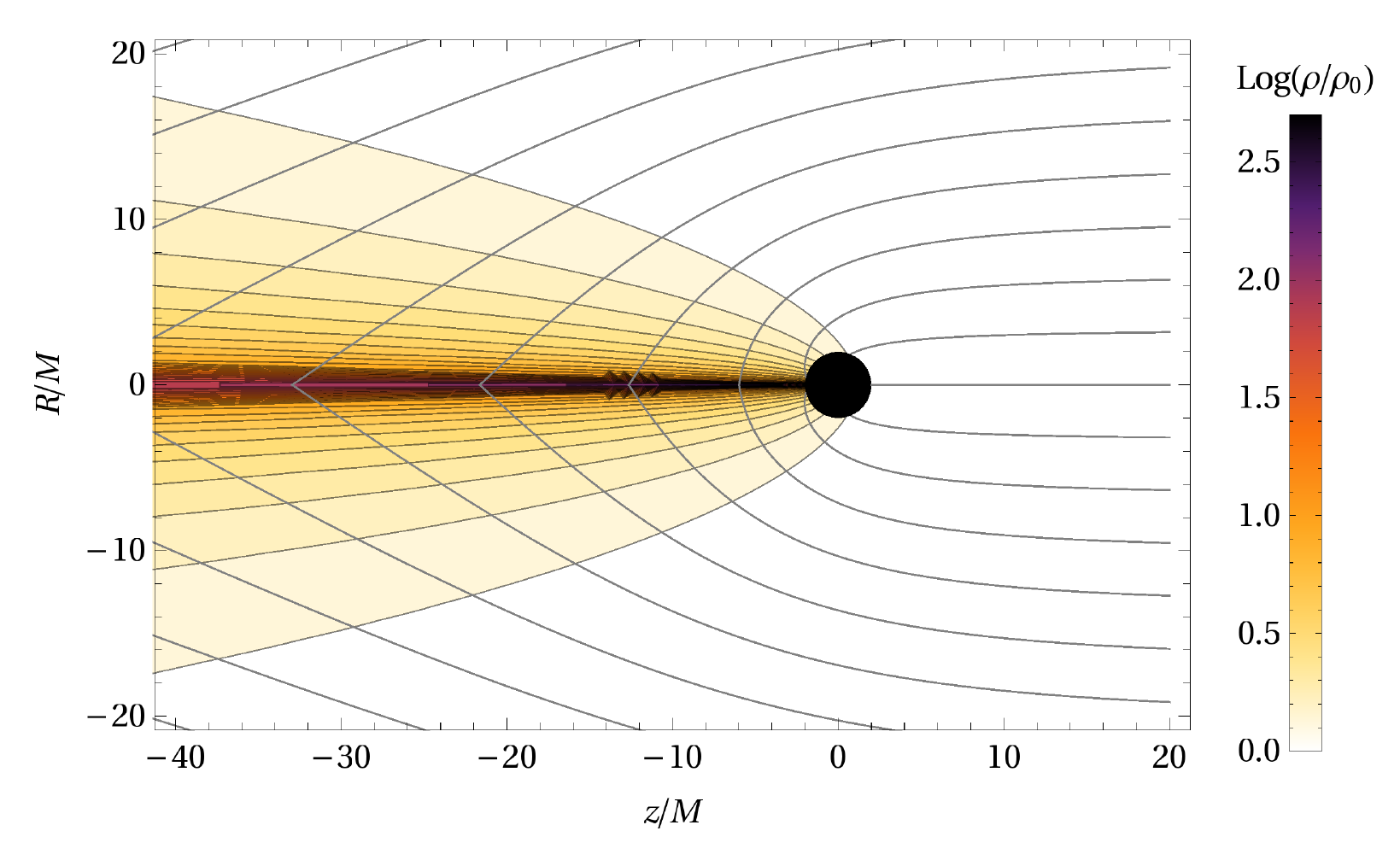}
\end{center}
\caption{Analytic model of a relativistic wind accreting at $v\8 = 0.5$ onto a 
Schwarzschild black hole. Accretion flow streamlines are represented as gray, 
solid lines, while colour contours correspond to the density field level set.}
\label{fig3}
\end{figure}

In Figure~\ref{fig3} we show an example of a wind accreting at \mbox{$v\8 = 
0.5$} onto a Schwarzschild black hole. The figure shows the flow streamlines as 
expressed analytically by \eqs{e3.9} and \eqref{e3.13} together with the 
density field as calculated numerically from \eq{e3.23}.

\begin{figure*}
\begin{center}
 \includegraphics[width=0.9\textwidth]{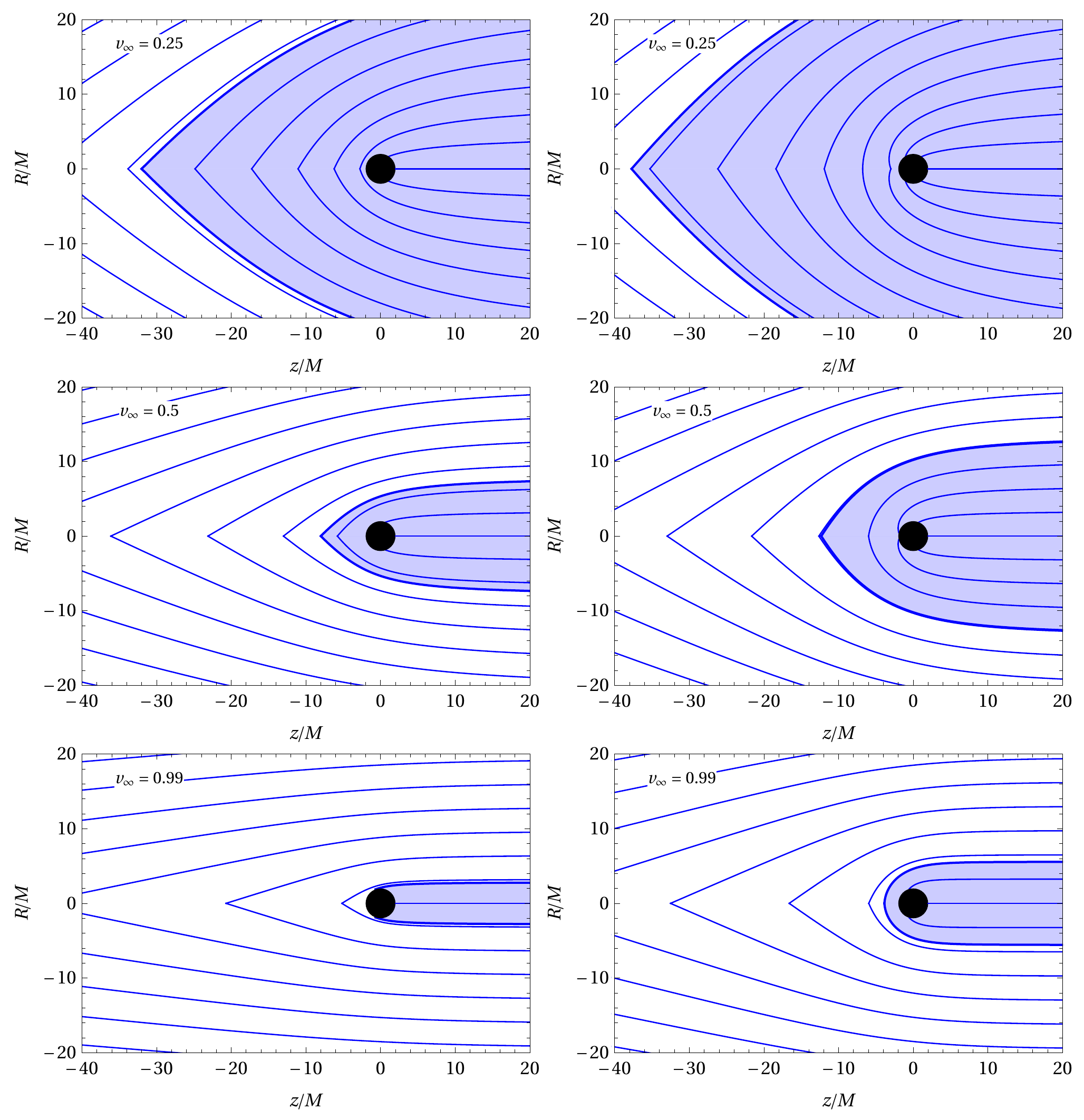}
\end{center}
\caption{Comparison of the ballistic streamlines as obtained in the BHL model 
(left-hand column) and in Schwarzschild spacetime (right-hand column). The 
shaded area in each panel corresponds to the accretion cylinder, i.e.~to those 
streamlines with impact parameter $b$ less than the critical value $b_c$ that 
end up accreting onto the central object. The wind speed at infinity $v\8$ is 
indicated at the top left of each panel. }
 \label{fig4}
\end{figure*}

In Figure~\ref{fig4} we compare the resulting streamlines in Schwarzschild 
spacetime with those coming from the Newtonian BHL model for different values 
of the asymptotic wind speed. From this figure it is clear that the accretion 
cylinder, and thus the total accretion rate, in the relativistic model is 
greater than the corresponding Newtonian value. In the next subsection we 
discuss this in further detail.

\subsection{Accretion rate}

Just as for the density field, the procedure described in Section~\ref{s2} to 
compute the total accretion rate in the Newtonian case (BHL model) is not as 
simple to implement for the relativistic problem. However, the logic behind 
this calculation remains the same: when the collision of mirror-reflected 
streamlines along the symmetry axis $(\theta = \pi)$ occurs, the flow's kinetic 
energy associated to the normal component of the velocity ($U^\theta$ in this 
case) is lost from the system as thermal energy and/or radiation. The remaining 
relativistic energy $\mathcal{E}'$ of a given streamline with impact parameter 
$b$ will determine whether the gas traveling along the streamline is accreted 
($\mathcal{E}'<1$) onto the central black hole or lost to infinity 
($\mathcal{E}' > 1$). The streamline that is left marginally bound with 
$\mathcal{E}' = 1$ is characterized by the critical impact parameter $b_c$. 
Using the normalization condition $U_\mu U^\mu = -1$, together with $U^\theta = 
0$ and \eq{e3.2}, it is simple to show that the condition $\mathcal{E}' = 1$ is 
equivalent to
\begin{equation}
 r(\pi)^3 + [2M - r(\pi) ]\,b_c^2 = 0.
 \label{e3.24}
\end{equation}
Unfortunately, after substituting $r(\pi)$ from either \eqs{e3.9} or 
\eqref{e3.13}, the resulting equation is highly non-linear in $b_c$ and it does 
not seem possible to solve it for $b_c$ explicitly.  Nonetheless, it is 
straightforward to solve \eq{e3.24} numerically using a root-finding algorithm. 
We have done this using the bisection method for 1000 values of $v\8$ uniformly 
distributed between 0.001 and 0.999 to a precision of 10$^{-8}$. Now, based on 
these numerically calculated values, we found the following fit that 
approximates $b_c$ to an accuracy better than $3\%$ for $v\8<0.999$ and to 
within $0.5\%$ for $v\8<0.98$:
\begin{equation}
b_c^\mathrm{fit} = \frac{2M}{v^2\8}\left(1 + c_1\,v\8 + c_2\,v^2\8 + 
c_3\,v^3\8\right),
\label{e3.25}
\end{equation}
with $c_1=0.081135$, $c_2=3.452826$ and $c_3= -1.758438$. 

All of the streamlines within the cylinder $b\le b_c$ contribute to the total 
accretion rate. According to the right-hand side of \eq{e3.22}, we can 
therefore express $\dot{M}$ as
\begin{equation}
\dot{M} = \pi\,b_c^2\,\rho\8 v\8 \gamma\8.
\label{e3.26}
\end{equation}

\begin{figure}
\centering
\includegraphics[width=0.45\textwidth]{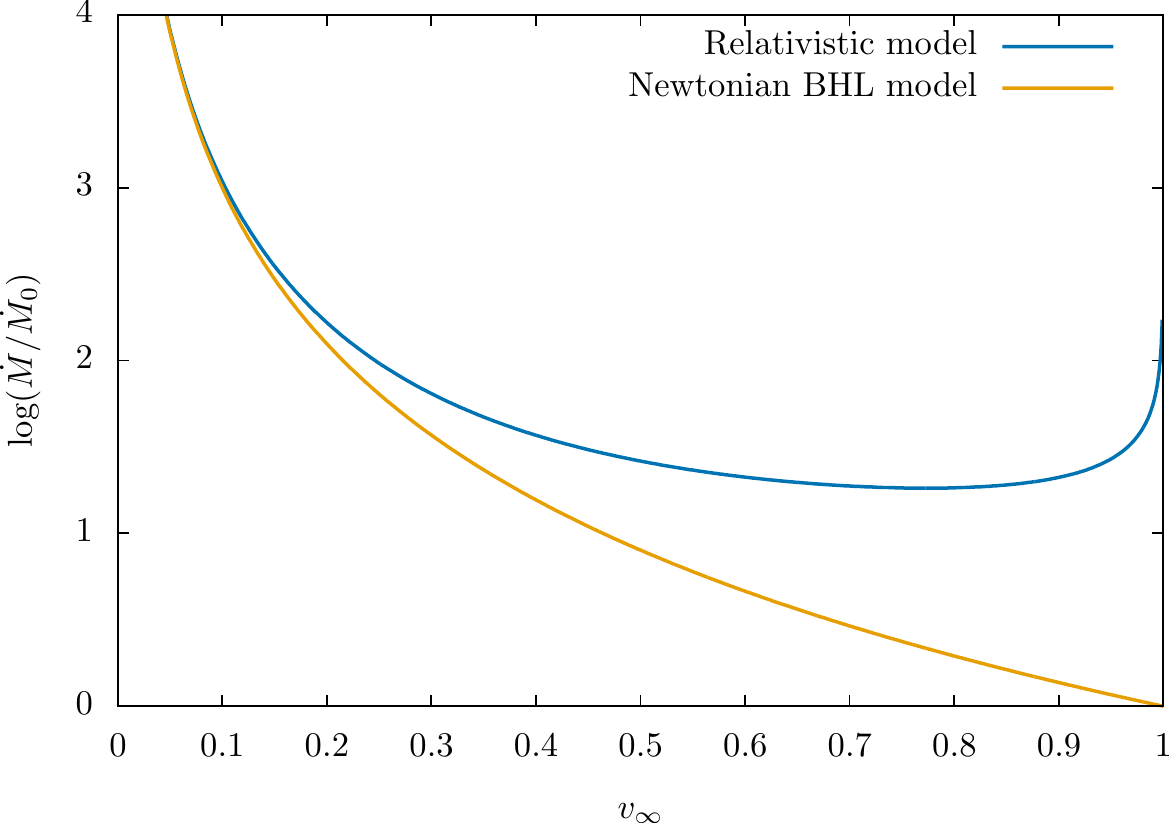}
\caption{Mass accretion rate as a function of the wind speed at infinity 
$v_{\infty}$. The blue line corresponds to the (fitted) relativistic model as 
given by Eq.~\ref{e3.25}, while the yellow line corresponds to the Newtonian 
BHL model (Eq.~\ref{e2.7}). In both cases the accretion rate is expressed in 
units of $\dot{M}_0 = 4\pi M^2 \rho\8 $.  }
\label{fig5}
\end{figure}

In Figure~\ref{fig5} we show the resulting accretion rate as a function of the 
wind speed at infinity and compare it to the Newtonian value found in the BHL 
model (Eq.~\ref{e2.7}). As can be seen from this figure, the relativistic 
accretion rate is greater than the corresponding BHL one for all values of 
$v\8$, with the difference being more than double for $v\8\gtrsim0.4$ and more 
than ten times larger for $v\8\gtrsim0.8$. As expected, the relativistic 
accretion rate converges to the Newtonian value in the non-relativistic limit, 
i.e~for $v\8\ll1$. Contrary to the Newtonian case in which the accretion rate 
was a monotonically decreasing function of $v\8$, the relativistic result has 
an inflection point at around $v\8\simeq 0.8$ and becomes arbitrarily large as 
\mbox{$v\8\rightarrow1$}. The culprit behind this behaviour is the special 
relativistic effect of Lorentz contraction that compresses the fluid volume 
elements along the direction of the wind and is represented by the Lorentz 
factor in \eq{e3.26}.

\setcounter{equation}{0} 
\section{Comparison with numerical simulations}
\label{s4}

In this section we compare the analytic wind accretion model presented in the 
previous section against numerical hydrodynamic simulations performed with 
\textit{aztekas}, a free GNU Public Licensed (GPL) code for solving any 
conservative set of equations, in particular, relativistic (with a non-trivial 
fixed metric) and non-relativistic hydrodynamic equations. Some aspects of 
\textit{aztekas}, together with numerical and convergence tests relevant to the 
present work are discussed in the Appendix \ref{sA}. Further details and tests 
of the \textit{aztekas} code will be presented elsewhere.

For all of the simulations discussed in this work, we considered a perfect 
fluid evolving on top a fixed Schwarzschild background metric described in 
terms of the horizon-penetrating, Kerr-Schild coordinates. 
Since the problem under study presents axial symmetry, all of the simulations 
were performed in 2D using polar coordinates $r$ and $\theta$. All of the 
results presented below were obtained after the numerical simulations had 
evolved in time from a uniform initial state condition until a relaxed 
stationary state was reached.

The relativistic hydrodynamic equations consist on the continuity equation 
(\ref{e3.19}) together with the local conservation of energy-momentum
\begin{equation}
\nabla_\mu T^{\mu \nu} = 0,
\label{e4.1}
\end{equation}
where we take the energy-momentum tensor $T^{\mu \nu}$ corresponding to a
perfect fluid~\citep{daufields}
\begin{equation}
T^{\mu \nu} = \rho\, h\, U^\mu U^\nu + p\,g^{\mu \nu},
\label{e4.2}
\end{equation}
with rest mass density $\rho$, pressure $p$, specific internal energy
$\epsilon$ and specific enthalpy $h = 1 + \epsilon + p/\rho$. Moreover,
we  adopt a Bondi-Wheeler equation of state~\citep{tooper1965} of the form
\begin{equation}
\epsilon = \frac{p}{\rho \left( \Gamma - 1 \right)},
\label{e4.3}
\end{equation}
where $\Gamma$ is the polytropic index.

In order to integrate numerically \eqs{e3.19} and~\eqref{e4.1} with
\textit{aztekas}, we recast them in a conservative form based on the 3+1
formalism~\citep[see e.g.][]{font2000,alcubierre}, as follows
\begin{equation}
 \frac{1}{\sqrt{-g}}\frac{\partial \left( \sqrt{\gamma} \, \mathbf{Q} 
\right)}{\partial t} +
\frac{1}{\sqrt{-g}} \frac{\partial \left( \sqrt{-g} \, \mathbf{F}^i
\right)}{\partial x^i} = \mathbf{S},
\label{e4.4}
\end{equation}
where $\gamma$ and $g$ are the determinants of the spatial 3-metric 
$\gamma_{ij}$ and the spacetime 4-metric $g_{\mu \nu}$, respectively. 
$\mathbf{Q}$ is the conservative variable vector, $\mathbf{F}^i = \left\lbrace 
\mathbf{F}^r, \mathbf{F}^\theta \right\rbrace$ are the fluxes along the $r$ and 
$\theta$ coordinates and $\mathbf{S}$ is the source vector. All of these 
quantities depend on the primitive variables $\mathbf{U}$. The functional form 
of these vectors are presented below:
 \begin{equation}
 \mathbf{U} = \left[ \rho,\ v_j,\ p \right]^T,
 \label{e4.5}
 \end{equation}
\begin{equation}
\mathbf{Q} = \left[ D,\ S_j,\ \tau\right] = \left[ \rho W,\ \rho\, h\, W^2 
v_j,\ \rho\, h\, W^2 - p - D \right]^T,
\label{e4.6}
\end{equation}

\begin{equation}
\begin{split}
\mathbf{F}^i = \bigg[ D\left( v^i - \frac{\beta^i}{\alpha} \right),\ &S_j 
\left( v^i - \frac{\beta^i}{\alpha} \right) + p\,\delta_j^i, \\
& \left.\tau\left( v^i - \frac{\beta^i}{\alpha} \right) + p\, v^i \right]^T,
\label{e4.7}
\end{split}
\end{equation}

\begin{equation}
\begin{split}
\mathbf{S} = \bigg[ 0,\ &T^{\mu \nu} \left( \frac{\partial g_{\nu j}}{\partial
x^\mu} - \Gamma_{\nu \mu}^\sigma g_{\sigma j} \right), \\
&\left. \alpha \left( 
T^{\mu 0} \frac{\partial \ln \alpha}{\partial x^\mu} - T^{\mu \nu}
\Gamma_{\nu \mu}^0 \right) \right]^T,
\label{e4.8}
\end{split}
\end{equation}
where $\Gamma_{\mu \nu}^\lambda$ are the usual Christoffel symbols, $v^i 
$ is the three-velocity as measured by local Eulerian observers, $W = 
\left( 1 - \gamma_{ij}v^i v^j \right)^{-1/2}$ is the associated Lorentz 
factor, and 
\begin{equation}
 \alpha = \left( 1 + \frac{2M}{r} \right)^{-1/2},
\label{e4.9}
\end{equation}
\begin{equation}
\beta^i = \left( \frac{2M}{r}\left( 1 + \frac{2M}{r} \right)^{-1},\ 0\ ,\ 0 
\right),
\label{e4.10}
\end{equation}
are the lapse function and the shift vector, respectively, and correspond to 
the 3+1 decomposition of Schwarzschild spacetime in Kerr-Schild coordinates.

For the numerical flux calculation, we use in \textit{aztekas} a high
resolution shock capturing method (HRSC) with an HLLE approximate Riemann
solver~\citep{hll1983}, combined with a monotonically centered (MC) second
order reconstructor at cell interfaces. For the time integration, we use a
second order Runge-Kutta method of lines in the total variation diminishing
(TVD) version~\citep{shu1988}. Finally, we adopt a constant time-step
defined through the CFL condition $\Delta t = C\,\min\left( \Delta r,
r\,\Delta \theta \right)$, with $C = 0.1$.

For the relativistic wind simulations, we ran two sets of simulations exploring
asymptotic wind speeds from $v\8=0.1$ to $0.9$ for two different polytropic
indices:  $\Gamma = 4/3$ and $5/3$. We took as numerical domain 
$[r_\mathrm{min},\
r_\mathrm{max}]\times[0,\pi]$, with $r_\mathrm{min} = 0.5\,r_\mathrm{acc}$,
$r_\mathrm{max} = 10\,r_\mathrm{acc}$ and
\begin{equation}
r_\mathrm{acc} = \frac{M}{v_\infty^2 + a_\infty^2 },
\label{e4.11}
\end{equation}
where $a_\infty$ is the asymptotic speed of sound. The radius 
$r_{\mathrm{acc}}$ is commonly used in the 
literature~\citep[cf.][]{font1999,cruzosorio2012}, as a good estimate for the 
extension of the numerical domain necessary for numerical 
convergence.\footnote{Note that with our choice of $r_\mathrm{min}$, 
simulations with $v\8 \ge 0.5$ are such that $r_\mathrm{min} < 2M$, which 
\textit{aztekas} can handle without any problem thanks to our choice of horizon 
penetrating coordinates. On the other hand, for those simulations with 
$r_\mathrm{min} > 2M$, we performed trial tests to make sure that the resulting 
steady-state accretion rate was unchanged independently of whether the black 
hole event horizon ($2M$) was part of the numerical domain or not.} Our choice 
of this numerical domain is based on the convergence and resolution tests 
discussed in detail in the Appendix~\ref{sA}.

For most of these simulations we used a fixed Mach number $\mathcal{M} = 
v\8/a\8 = 5$ in order to ensure that we were considering the same supersonic 
conditions in all cases. As a sanity check, we also considered $\mathcal{M} = 
10$ for a reduced number of cases. We used a uniform grid of $400\times400$ 
cells except for the cases $v_\infty = 0.1$, $0.8$ and $0.9$, where we had to 
use a larger grid of $1000\times1000$ in order to find converged solutions.

As external boundary condition at the sphere $r=r_\mathrm{max}$, we enforced a 
constant, uniform inflow from the northern hemisphere $\theta\in[0,\pi/2)$ and 
free outflow from the southern one $\theta\in[\pi/2,\pi]$. As internal boundary 
condition, we set free outflow across the sphere at $r=r_\mathrm{min}$. The 
incoming wind at the external boundary has uniform thermodynamical variables 
$\rho\8$ and $p\8$. We set $\rho\8=10^{-10}$ in arbitrary units and take $p\8$ 
consistent with the chosen Mach number $\mathcal{M}$ and the equation of state, 
as in~\citet{cruzosorio2012}:
\begin{equation}
p_\infty = \frac{a_{\infty}^2 \rho_\infty (\Gamma - 1)}{\Gamma (\Gamma - 1 )- 
a_{\infty}^2 \Gamma }.
\label{e4.12}
\end{equation}

For the velocity field, we set an incoming wind with a constant velocity 
$\sqrt{v_i\,v^i} = v_\infty$, and components:
\begin{equation}
v_r = -\sqrt{g_{rr}}\,v_\infty \cos \theta,
\label{e4.13}
 \end{equation}
\begin{equation}
v_\theta = \sqrt{g_{\theta \theta}}\,v_\infty \sin \theta,
\label{e4.14}
\end{equation}
where $g_{rr}$ and $g_{\theta \theta}$ are the radial and polar components of 
the metric respectively. As for the initial conditions of the simulation, we 
set them equal to the constant boundary values over all the numerical domain.

The simulations were left to run until a stationary state was reached. This was 
monitored by keeping track of the mass accretion rate, which was computed on 
the fly by integrating the relativistic radial mass flux across a control 
sphere of radius $r$ according to \citep{petrich89}
\begin{equation}
\dot{M} = 2\pi \int_{0}^{\pi} D \left(v^r - \frac{\beta^r}{\alpha} \right) \, 
r^2 \sin \theta \mathrm{d} \theta.
\label{e4.15}
\end{equation}
It is important to remark that, once the stationary state is reached across all
the numerical domain, the mass accretion rate as calculated from \eq{e4.15} has
to have a constant value (to within numerical precision) independently of where
the control radius $r$ is located (see Figure~\ref{figA3}).

\begin{figure}
\centering
\includegraphics[width=0.45\textwidth]{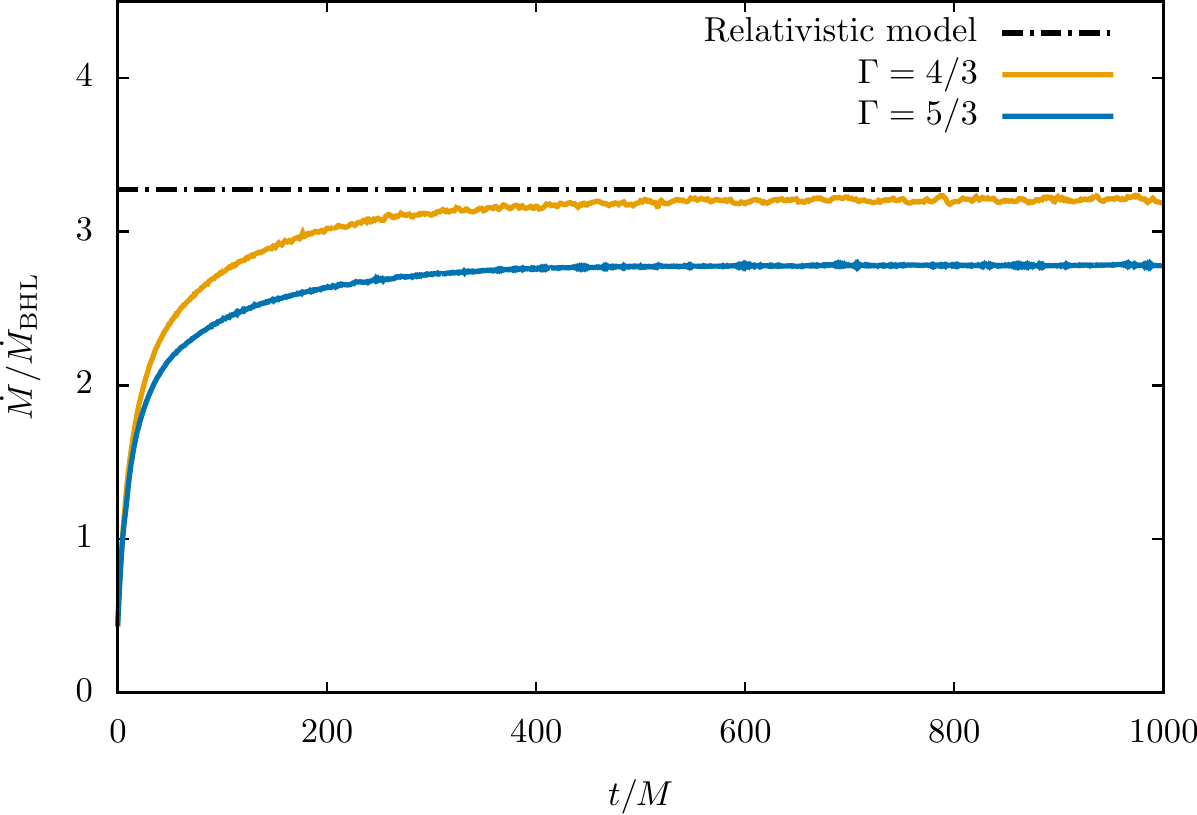}
\caption{Mass accretion rate as a function of time for two wind simulations 
with asymptotic velocity $v\8 = 0.5$ and  $\Gamma=4/3,\  5/3$. As a reference, 
the horizontal dashed line shows the corresponding value from the relativistic 
model (Eq.~\ref{e3.26}). The stationary state is reached after $t \approx 600 
M$. The mass accretion rate has been scaled using the BHL result 
$\dot{M}_\mathrm{BHL} = 4\pi M^2 \rho\8 / v_\infty^3$ (Eq.~\ref{e2.7}).}
\label{fig6}
\end{figure}

The typical simulation time at which steady state was reached depends on the 
wind velocity at infinity and the domain extension roughly as $t_c \approx 
r_{\mathrm{acc}}/v\8$. As an example, in Figure~\ref{fig6} we show the mass 
accretion rate as a function of time for $v\8 = 0.5$ and $\Gamma = 4/3,\ 5/3$, 
measuring it at the event horizon $r = 2M$.

\begin{figure*}
\centering
\includegraphics[width=0.49\textwidth]{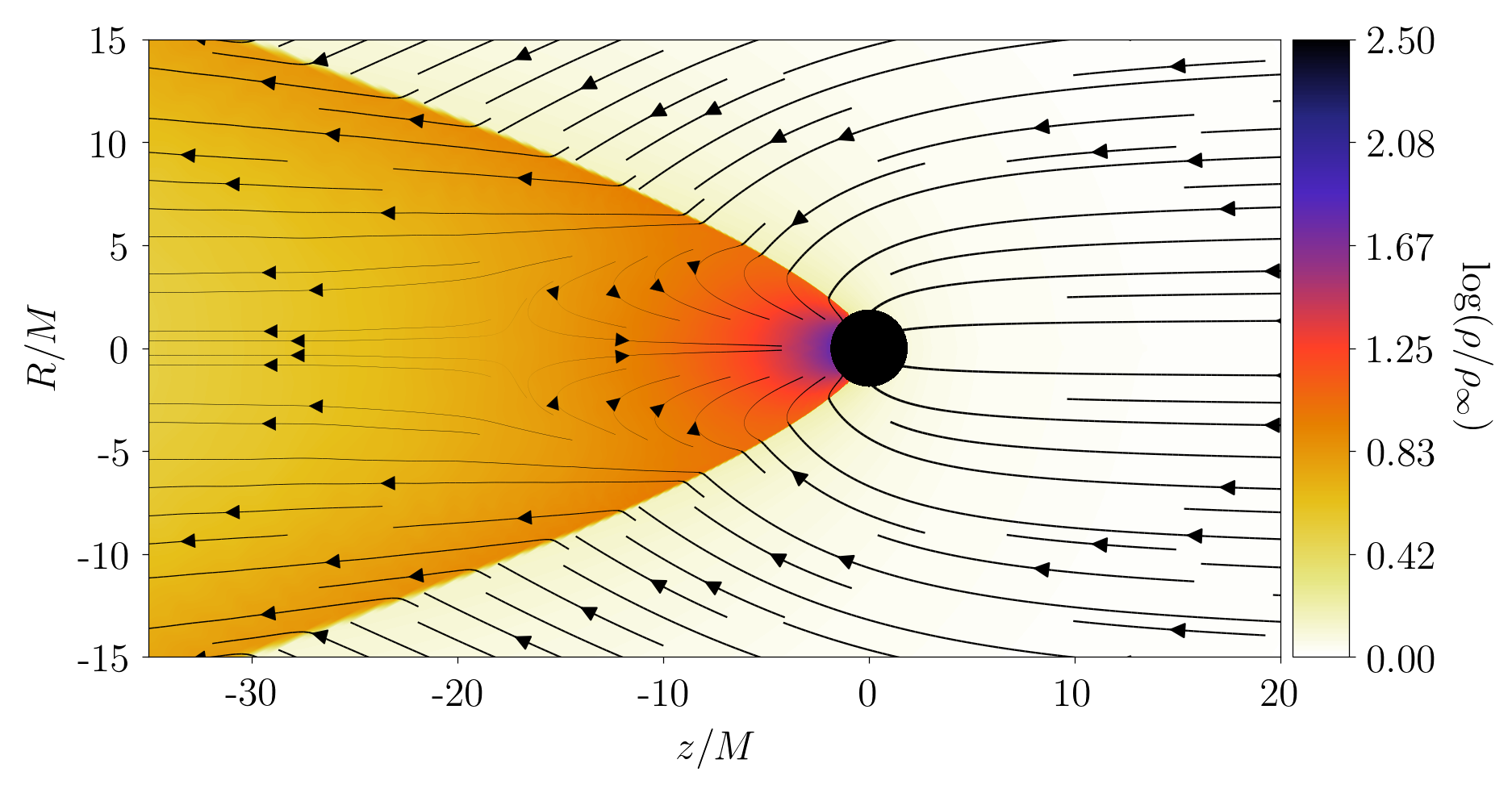}
\includegraphics[width=0.49\textwidth]{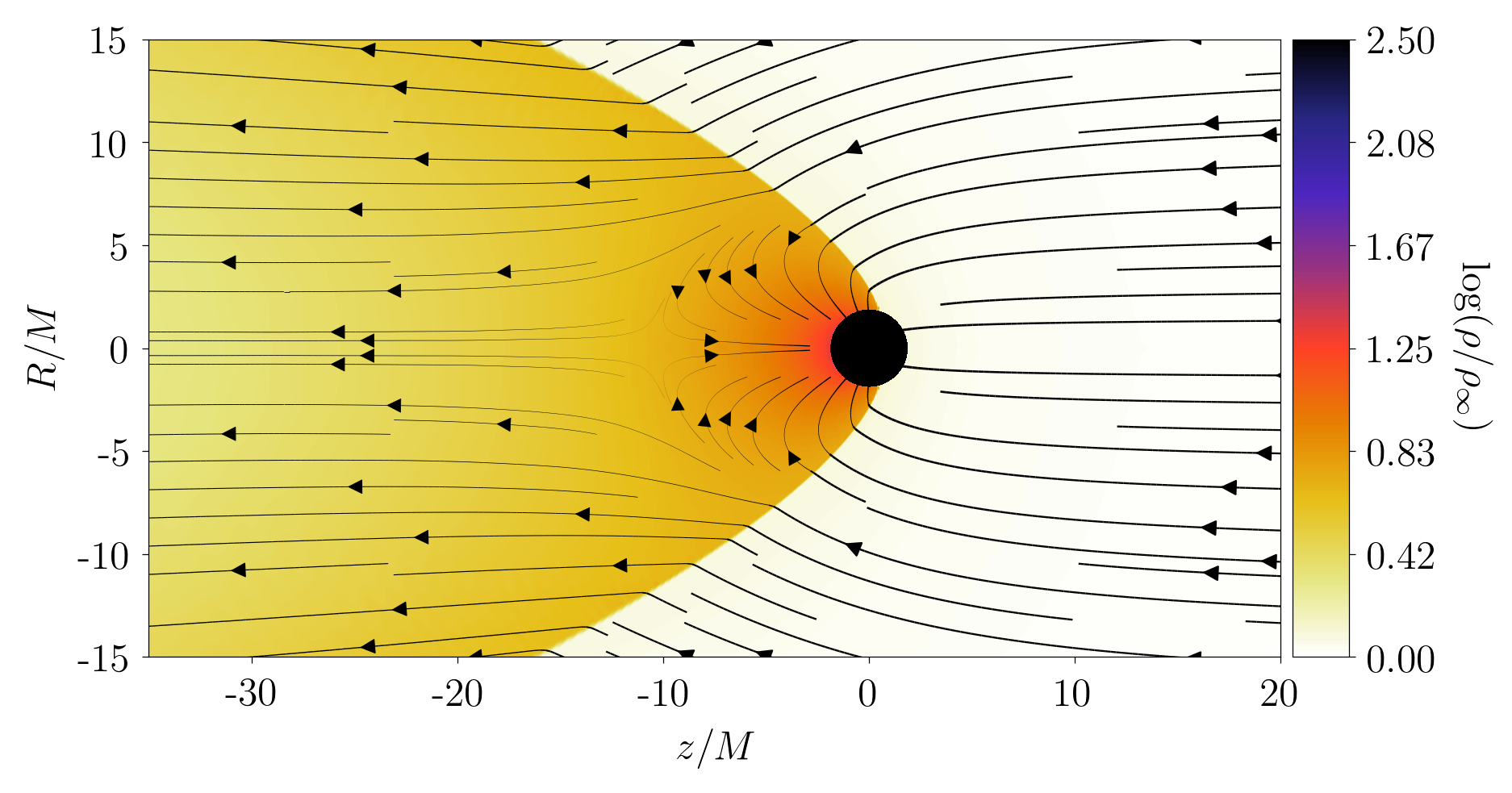}
\caption{Numerical simulations performed with \textit{aztekas}  of a 
relativistic wind  accreting onto a Schwarzschild black hole. The chosen 
polytropic index is $\Gamma = 4/3$ for the left side panel and $\Gamma = 5/3$ 
for the right side one, while the asymptotic wind velocity is $v\8=0.5$. The 
plot shows the stationary-state streamlines and in colour isocontour levels of 
the corresponding density field. The width of the streamlines is proportional 
to the velocity magnitude $v = \sqrt{v_i\,v^i}$. }
\label{fig7}
\end{figure*}

In Figure~\ref{fig7}, we show the steady-state of the density field and 
streamlines of a wind accretion flow with $v_\infty = 0.5$ and polytropic index 
$\Gamma=4/3$ on the left side panel and  $\Gamma=5/3$ on the right side panel. 
As expected for a supersonic flow, a bow shock is formed downstream around the 
accretor, with a smaller shock cone for \mbox{$\Gamma=4/3$} than for 
$\Gamma=5/3$. 

\begin{figure}
\centering
\includegraphics[width=0.45\textwidth]{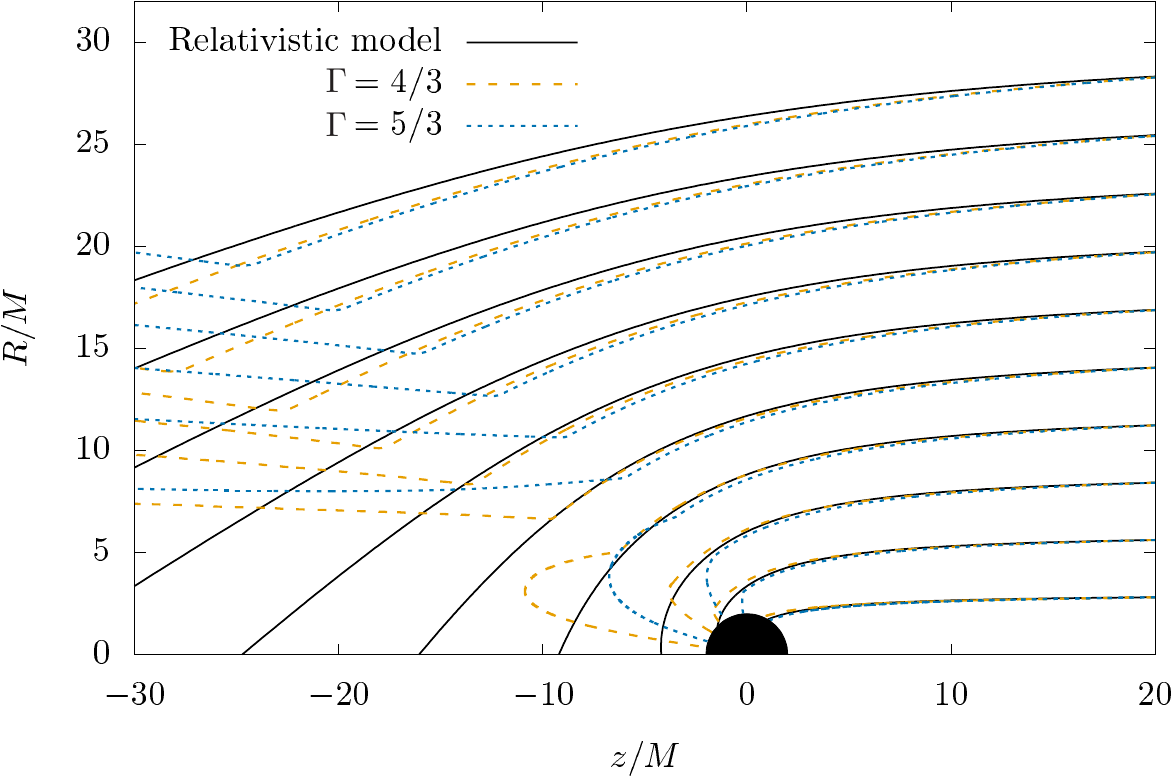}
\caption{Comparison of the streamlines of the analytic model against the 
corresponding ones extracted from the numerical simulations for 
\mbox{$v\8=0.5$}, $\mathcal{M} = 5$ and  $\Gamma = 4/3,\ 5/3$. }
\label{fig8}
\end{figure}

In Figure~\ref{fig8}, we compare the streamlines as obtained from the two 
simulations discussed in the previous paragraph against the ones from the 
analytic model. As can be seen from this figure, the ballistic approximation 
provides a qualitatively good  description of the resulting streamlines 
upstream of the flow and in the region outside the bow shock, while inside the 
shock cone the streamline behavior is notably different. 

We can also notice from both Figure~\ref{fig6} and Figure~\ref{fig8}, that 
there is a better agreement between the analytic model and the $\Gamma=4/3$ 
case than for the $\Gamma=5/3$ one. This is related to the fact that the 
relative degree of incompressibility or stiffness of a polytropic fluid is 
directly proportional to the adiabatic index and, thus, a $\Gamma=5/3$ fluid 
resists more effectively the compression due to the gravitational field of the 
central object (geodesic focusing) than a fluid with $\Gamma=4/3$.

\begin{figure}
\centering
\includegraphics[width=0.45\textwidth]{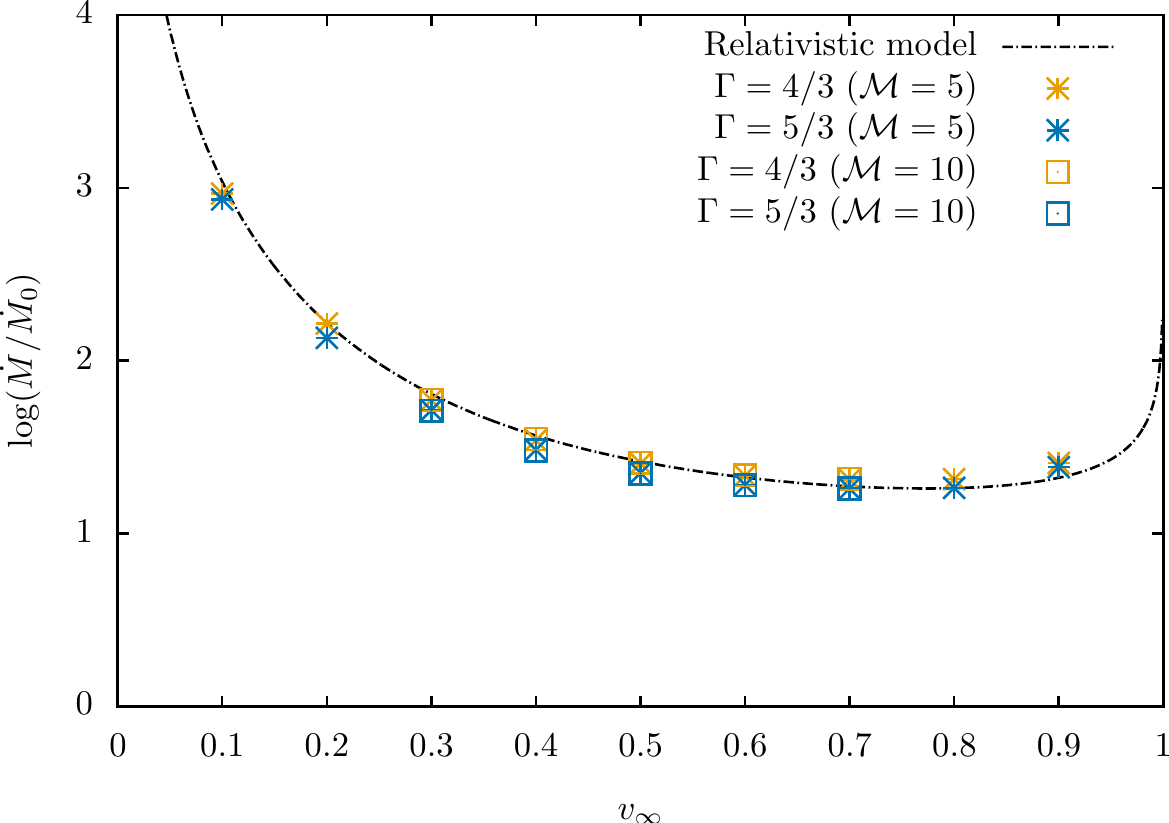}
\caption{Comparison of the mass accretion rate as obtained from the 
relativistic model (Eq.~\ref{e3.26}) versus the numerical values obtained 
from the \textit{aztekas} simulations. The colour marks show the results 
of the 28 simulations discussed in this work for asymptotic wind speeds 
$v\8 = 0.1,\,0.2,\ldots,0.9$, two polytropic indices 
$\Gamma = 4/3,\,5/3$, and two Mach numbers $\mathcal{M}=5,\,10$. The mass 
accretion rate is expressed in units of $\dot{M}_0 = 4\pi M^2\rho\8$.}
\label{fig9}
\end{figure}

In order to compare the numerical mass accretion rate with the ballistic model 
(Eq.~\ref{e3.24}), we computed the mean value of \eq{e4.15} across all the 
radial domain. In Figure~\ref{fig9} we show the comparison between the analytic 
model presented in the previous section and the simulations for both $\Gamma = 
4/3$ and $5/3$. 

Moreover, we also explored the dependence of the mass accretion rate on the 
Mach number by performing 10 additional simulations with  $\mathcal{M} = 10$. 
In Figure~\ref{fig9}, the square marks show the mass accretion rate for these 
simulations. Note that these results overlay with the $\mathcal{M} = 5$ cases 
(the relative difference between both results is less than $0.7\%$). The 
resulting bow shocks are also very similar in both cases, with slightly smaller 
cones for the $\mathcal{M} = 10$ case.

As can be seen in Figure~\ref{fig9}, the resulting accretion rate from the 
numerical simulations is consistently slightly larger for the \mbox{$\Gamma = 
4/3$} simulations than for the $\Gamma =5/3$ ones. Moreover, a good agreement 
is found between the accretion rate as predicted by the relativistic model and 
the one obtained from the \textit{aztekas} simulations, with a relative error 
less than 10\% for all of the 28 simulations presented in this work.

\section{Summary}
\label{s5}

We have presented a full relativistic, analytic model of a supersonic wind 
accreting onto a Schwarzschild black hole. In addition to the assumptions of 
stationarity and axisymmetry, the model is based on the ballistic approximation 
in which the streamlines of the accretion flow correspond to geodesic 
trajectories of a Schwarzschild spacetime. Following the methodology presented 
in \cite{tejeda2}, the streamlines of the model were described analytically in 
terms of Jacobi elliptic functions. The density field of the resulting 
accretion flow and the corresponding accretion rate were calculated using 
simple numerical schemes. 

The model presented in this paper constitutes the relativistic generalisation 
of the Newtonian wind accretion model by Bondi-Hoyle-Lyttleton (BHL). 
Naturally, the relativistic model recovers the BHL model in the 
non-relativistic limit $v\8 \ll \cc$. The enhanced gravitational field of the 
accreting object in general relativity, together with the special relativistic 
effect of Lorentz contraction, contribute to a larger accretion rate as 
compared to that of the BHL model (see Figure~\ref{fig5}). This difference 
becomes substantial (by a factor of 10-100) for asymptotic wind speeds $v\8$ 
close to the speed of light c. Although these large velocities are not expected 
to be common in astrophysical settings, they can appear in extreme cases such 
as the velocity kick imparted onto a newborn black hole following an 
asymmetrical supernova explosion \citep{janka13} or after the merger of two 
rotating black holes \citep{gerosa16}.

We have compared the new relativistic model against numerical simulations 
performed with the \textit{aztekas} code. This code solves numerically the full 
hydrodynamic evolution of a perfect fluid in Schwarzschild spacetime 
starting off from a uniform condition until a stationary state is reached. We 
have used two different polytropic equations of state (\mbox{$\Gamma = 4/3$} 
and \mbox{$\Gamma = 5/3$}) and asymptotic wind speeds from \mbox{$v\8 = 
0.1\,\cc$} to \mbox{$v\8= 0.9\,\cc$}. We have considered two different values 
for the asymptotic Mach number, $\mathcal{M} = 5$ and 10, finding virtually no 
difference between these two values. We have found a good agreement (to within 
10\%) between the accretion rate predicted by the relativistic model and the 
\textit{aztekas} simulations (see Figure~\ref{fig9}). As expected for these 
supersonic flows, the ballistic streamlines of the analytic model agree quite 
well with the resulting streamlines of the numerical simulations in the upwind 
region outside the bow shock. 

\section{Acknowledgements}

We thank Sergio Mendoza, Alejandro Cruz-Osorio, Olivier Sarbach and Francisco 
S.~Guzm\'an for useful discussions and comments on the manuscript. We also thank 
the anonymous referee for helpful suggestions and remarks. This work was 
supported by DGAPA-UNAM (IN112616 and IN112019) and CONACyT (CB-2014-01 
No.~240512; No.~290941; No.~291113) grants.

\bibliographystyle{mn2e}
\bibliography{references}

\appendix

\setcounter{section}{0}
\setcounter{equation}{0}

\renewcommand{\theequation}{A.\arabic{equation}}

\section{ Validation of the numerical hydrodynamic code {\it aztekas}}
\label{sA}
In this Appendix we present several numerical tests intended to validate
our use of {\it aztekas} in this work.

\subsection{Spherical accretion}

In order to test {\it aztekas} in the general relativistic hydrodynamic regime,
it is important to compare with a benchmark solution. For this, we employ the
analytic model of spherical accretion developed by \citet{michel72}.

\begin{figure}
\centering
\includegraphics[width=0.45\textwidth]{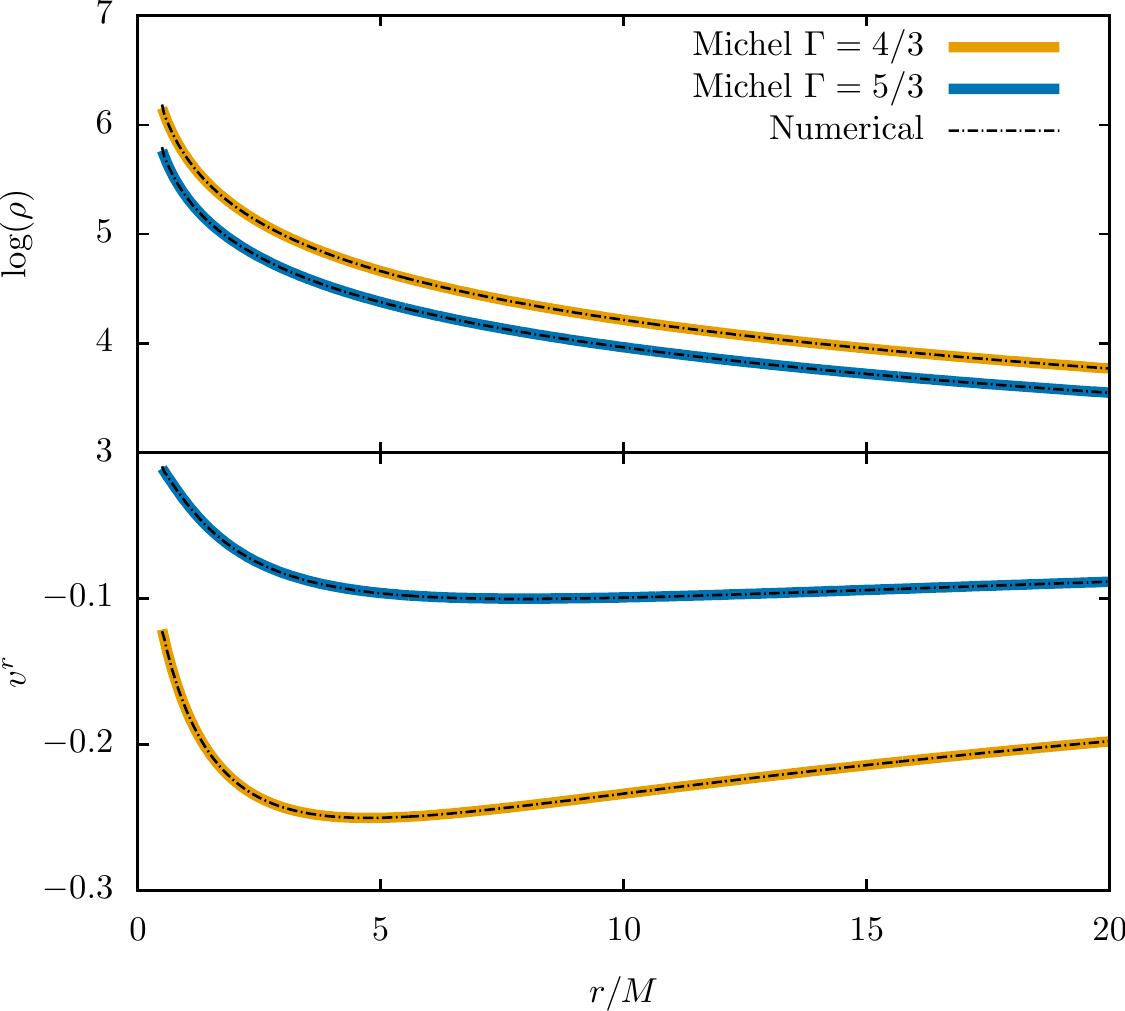}
\caption{Comparison of the numerical hydrodynamic simulations performed with 
\textit{aztekas}  against the relativistic spherical accretion model of 
\citet{michel72}. The top panel shows the density as a function of radius while 
the bottom panel shows the radial velocity for two different values of the 
polytropic index $\Gamma =4/3,\ 5/3$ and an asymptotic sound speed of $a\8 = 
0.01$.}  
\label{figA1}
\end{figure}

In Figure~\ref{figA1}, we compare the outcome of numerical simulations
performed with {\it aztekas} against Michel's analytic solution for the same
values of polytropic index used in the wind accretion problem ($\Gamma = 4/3,
5/3$) and for an asymptotic sound speed of $a\8 = 0.01$.

For these simulations, we took a 2D spherical axisymmetric grid of $400 \times
400$ uniformly distributed radial bins $r\in\left[ 0.5M,20M \right]$ and polar
bins $\theta \in \left[ 0,\pi/2 \right]$. We set the boundary conditions at
$r=20M$ by imposing Michel's solution there. As initial conditions we populate
the entire numerical domain with the boundary constant value and let the system
evolve until a stationary regime is reached for $t\gtrsim 100M$. As can be seen
from Figure~\ref{figA1}, an excellent agreement is found between Michel's
analytic solution and the {\it aztekas} simulation results. We compute the mass
accretion rate and the relative error between both solutions is below 0.2\%.

\begin{figure}
\centering
\includegraphics[width=0.45\textwidth]{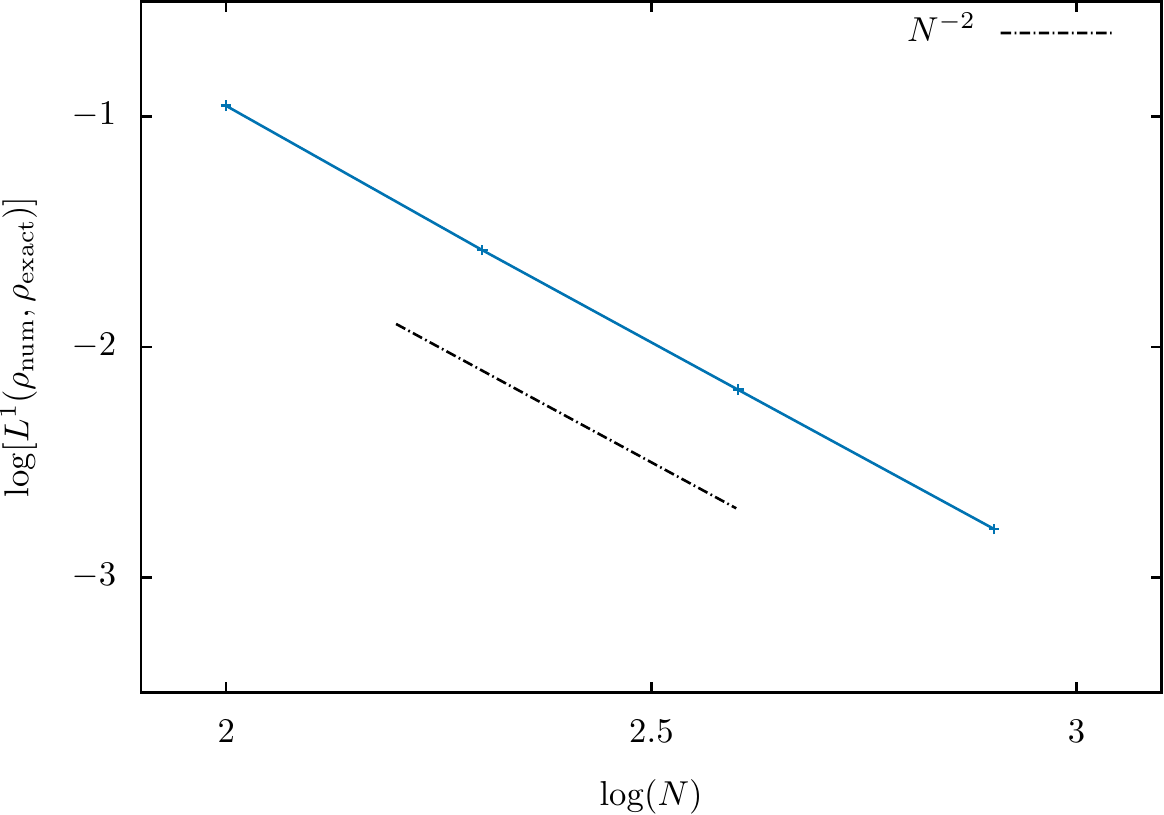}
\caption{$L^1$ norm of the density error for the Michel test, using resolutions 
$N_r = 100,\, 200,\, 400\, \mathrm{and}\, 800$. The dashed line shows the slope 
that corresponds to a second order convergence rate.}
\label{figA2}
\end{figure}

Also using this benchmark solution, we looked at the convergence rate of {\it 
aztekas} by computing the $L^1$ norm of the density error for different numerical 
resolutions $N_r$,\footnote{Since no angular dependence is found for this 
spherically symmetric test, we only varied the radial resolution while keeping a 
constant $N_\theta = 400$.} i.e.
\begin{equation}
L^1(\rho_{\mathrm{num}},\rho_{\mathrm{exact}}) = \frac{1}{N_r} \sum_{i=1}^{N_r} 
|\rho_{\mathrm{num}}(r_i) - 
\rho_{\mathrm{exact}}(r_i)|,
\label{eA1}
\end{equation}
where $\rho_\mathrm{num}$  and $\rho_\mathrm{exact}$ are the numerical and exact 
values of the density, respectively. In Figure~\ref{figA2}, we show 
the result of this test from where we obtain a convergence rate consistent with 
a second order, which is to be expected for smooth solutions and for our use of 
the MC reconstructor.

\subsection{Dependence on domain extension and resolution}

We also want to make sure that the numerical solutions obtained with {\it 
aztekas} have converged to physical values and that the results are independent 
from both numerical resolution and domain extension.

\begin{figure}
\centering
\includegraphics[width=0.45\textwidth]{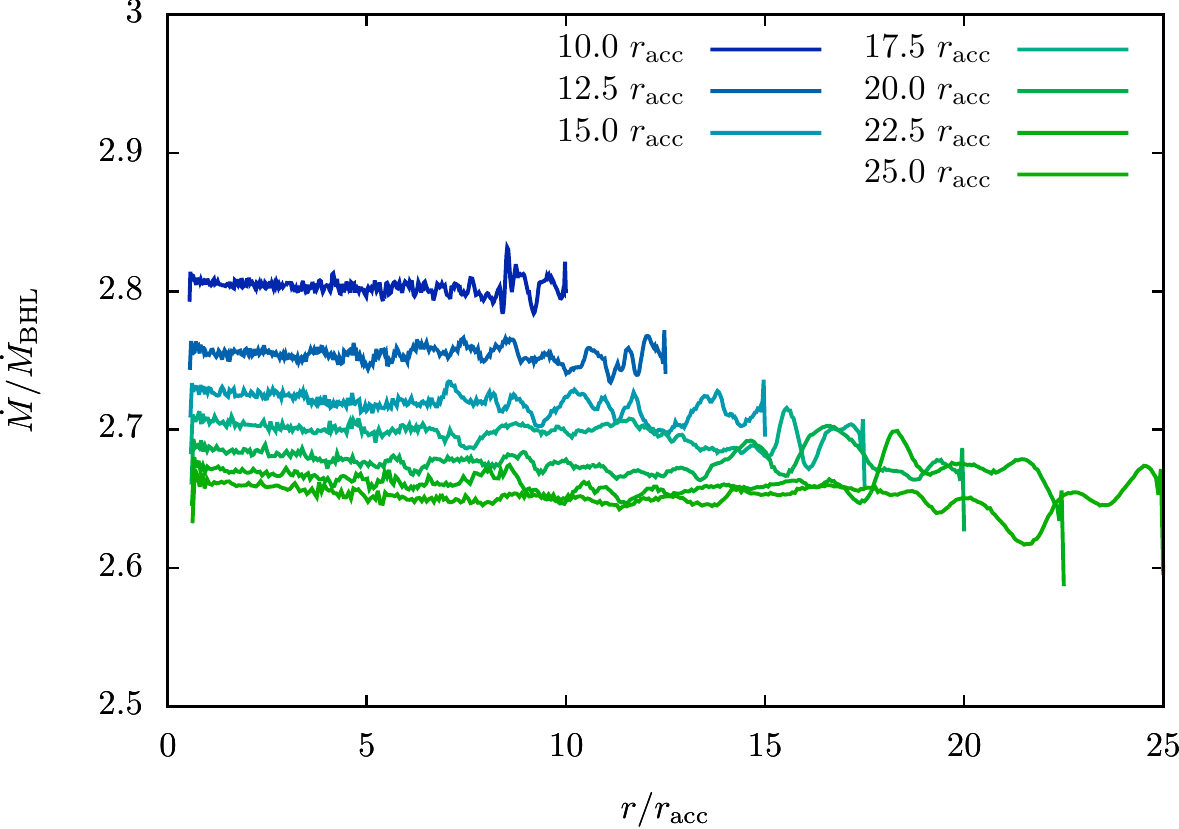}
\caption{Comparison of the mass accretion rate for different domain extensions. 
In all cases $v\8 = 0.5$, $\Gamma = 5/3$, and the simulation time is $t= 1000 
M$.}
\label{figA3}
\end{figure}

Due to the finite extension of the numerical domain $r_{\mathrm{max}}$, the 
total accretion rate calculated from the numerical simulations shows a small 
dependence on $r_{\mathrm{max}}$ that gets weaker as larger domain extensions 
are considered. For example, in Figure~\ref{figA3} we show the resulting 
mass accretion rate for simulations with $v\8=0.5$, $\Gamma = 5/3$ and seven 
different domain extensions, at $t=1000 M$.  As can be seen from this figure, 
our choice of a domain extension of $r_{\mathrm{max}} = 10\, r_{\mathrm{acc}}$ 
leads to an overestimation of the mass accretion rate of the order of $\sim 
5\%$, which is an acceptable margin of error for us in this work. We expect the 
rest of the simulations to behave in a similar way.

\begin{figure}
\centering
\includegraphics[width=0.45\textwidth]{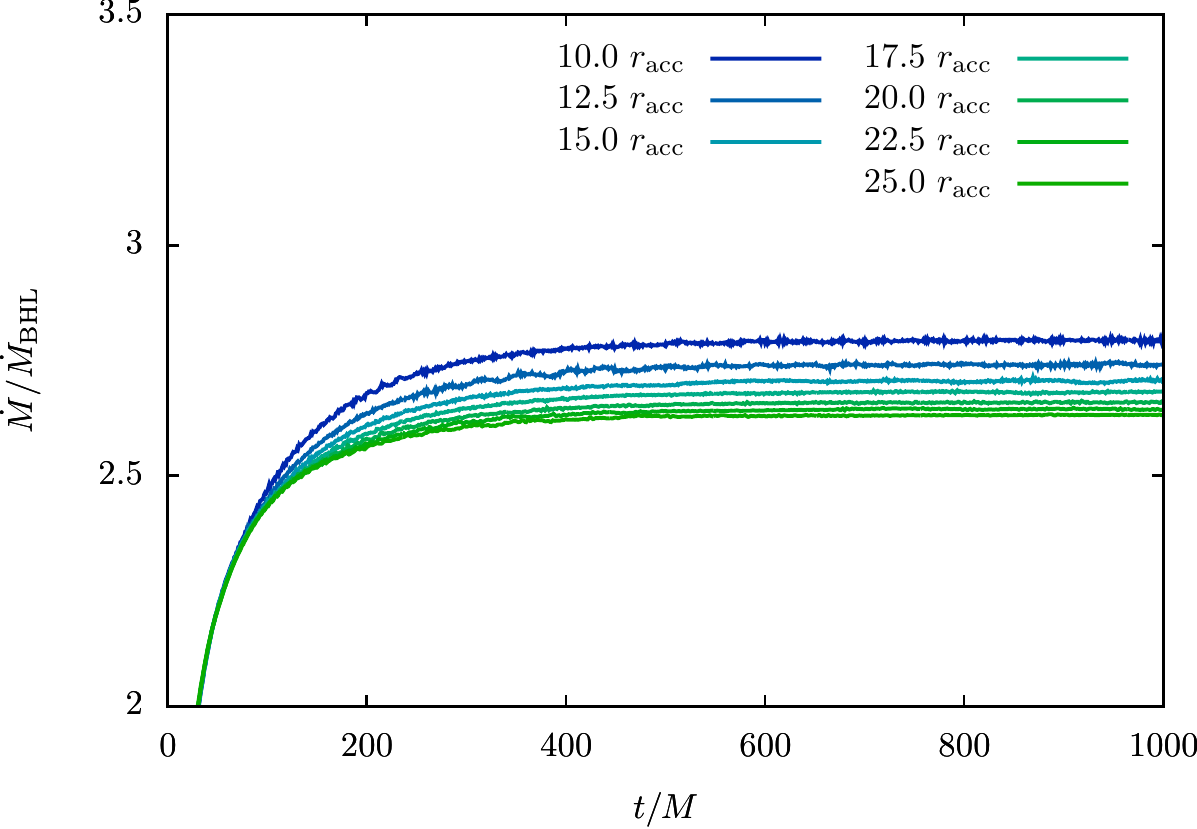}
\caption{Same as in Figure~\ref{figA3}, but now measuring the mass 
accretion rate across $r = 2M$ as a function of time.}
\label{figA4}
\end{figure}

Similarly, we also looked at this same domain extension dependence but now by 
monitoring the mass accretion rate across a fixed radius (in this case across 
the event horizon at $r=2M$) as a function of time. As can be seen from 
Figure~\ref{figA4}, the difference between a domain extension of $10 
\, r_{\mathrm{acc}}$ and of $25 \, r_{\mathrm{acc}}$ maintains the same margin 
error of $\sim 5\%$ along time.

\begin{figure}
\centering
\includegraphics[width=0.45\textwidth]{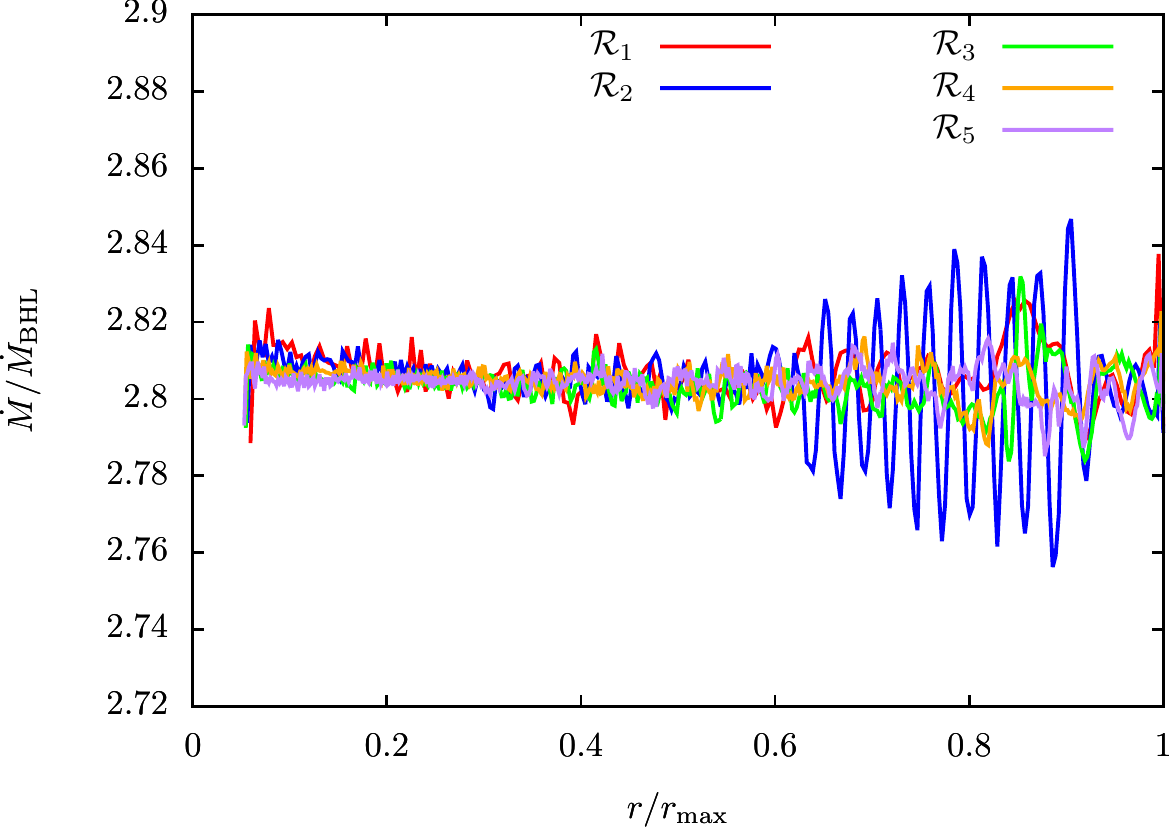}
\caption{Comparison between the mass accretion rate for different resolutions 
$\mathcal{R}_1 = 200\times200$, $\mathcal{R}_2 = 300\times300$, $\mathcal{R}_3 = 
400\times400$, $\mathcal{R}_4 = 500\times500$ and $\mathcal{R}_5 = 
600\times600$, for $v_\infty = 0.5$, $\Gamma = 5/3$ and $t=1000M$.}
\label{figA5}
\end{figure}

In what regards the numerical resolution, we found that the grid size does not 
affect as directly the value of the resulting accretion rate. However, it does 
contribute to the smoothness of the solution, with larger resolutions leading to 
less numerical noise. In Figure~\ref{figA5} we show an example of this 
again for the case $v\8=0.5$, $\Gamma = 5/3$ and five different numerical 
resolutions: $\mathcal{R}_1 = 200\times200$, $\mathcal{R}_2 = 300\times300$, 
$\mathcal{R}_3 = 400\times400$, $\mathcal{R}_4 = 500\times500$ and 
$\mathcal{R}_5 = 600\times600$. As discussed in the Section~\ref{s4}, for most 
of the simulations presented in this work we settled for $\mathcal{R}_3$ as a 
good compromise between accuracy and performance.

\begin{figure}
\centering
\includegraphics[width=0.45\textwidth]{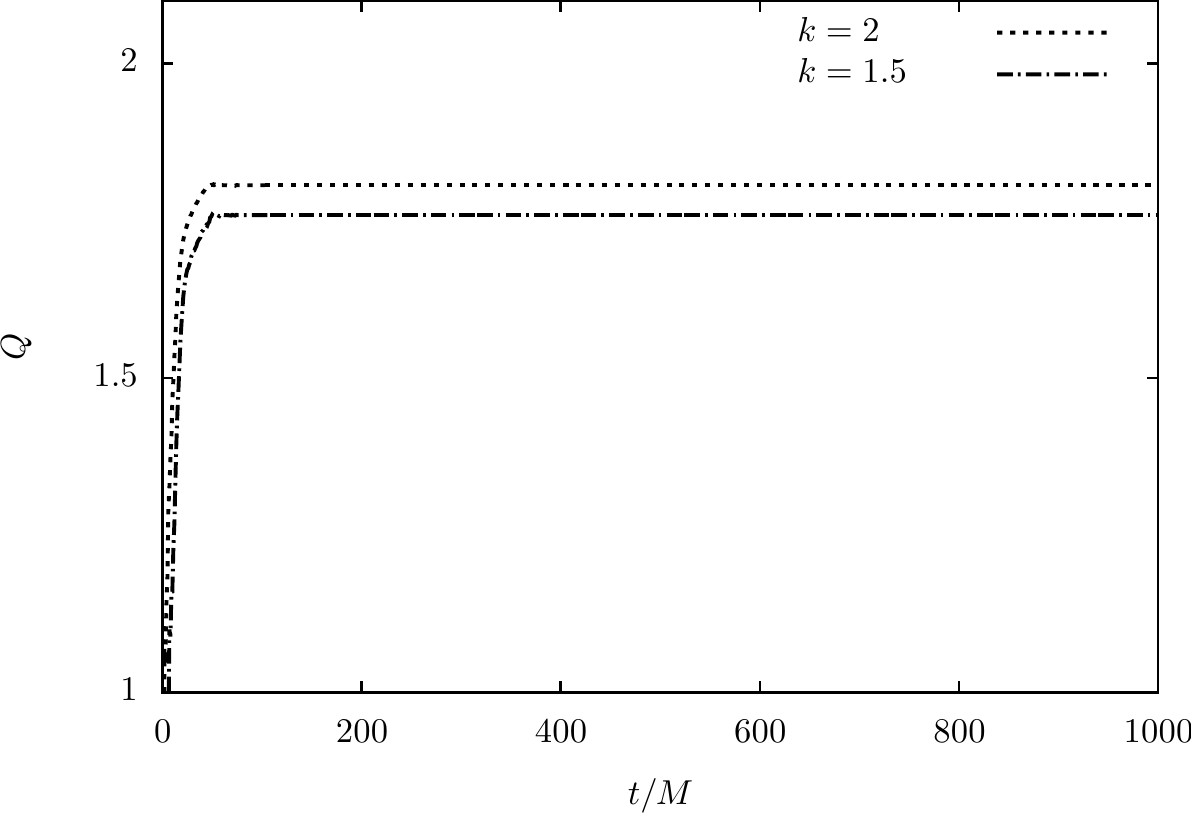}
\caption{Self-convergence of the wind simulations (see Eq.~\ref{eA2}) for two 
sets of simulations with a ratio of $k=1.5$ and $2$ between successive 
resolutions. In both cases the convergence rate $Q$ stays between 1.5 and 2 
in the steady state region, as required for this test due to the presence of 
shocks.}
\label{figA6}
\end{figure}

Finally, in order to estimate the convergence rate of the simulations, and given 
that there is no analytical solution for the full hydrodynamic wind accretion, 
we follow \citet{LG13} and measure the self-convergence rate using different 
resolutions. To this purpose, we ran two sets of three simulations each, with 
resolutions 
$$\mathcal{R}_1= 200\times200,\, \mathcal{R}_2=300\times300 \text{ and } 
\mathcal{R}_3=450\times450,$$
and
$$\mathcal{R}_1= 200\times 200,\, \mathcal{R}_2=400\times400 \text{ and } 
\mathcal{R}_3=800\times800.$$ Note that, for each set separately, there is a 
factor of $k=1.5$ and $k=2$ between successive resolutions. With the simulation 
results at hand, we compute the self-convergence factor $Q$ according to
\begin{equation}
k^Q = \frac{L^1(\rho_1,\rho_2)}{L^1(\rho_2,\rho_3)},
\label{eA2}
\end{equation}
with $L^1(\rho_i,\rho_j)$ as defined in \eq{eA1} and where $\rho_1$, 
$\rho_2$ and $\rho_3$ are the densities along the accretion axis $(\theta = 
\pi)$ for each of the employed resolutions. The results for this test are shown 
in Figure~\ref{figA6}. As can be seen from this figure, for both set of 
resolutions we find an order of convergence between 1.5 and 2 in the steady 
state region, which is to be expected for this kind of algorithms due to the 
presence of shocks.

\label{lastpage}

\end{document}